\documentclass[manuscript]{aastex}
\begin{document}
\title{The White Dwarfs within 20 Parsecs of the Sun:
Kinematics and Statistics}

\author{Edward M. Sion}
\affil{Dept. of Astronomy \& Astrophysics,
Villanova University,
Villanova, PA 19085, 
e-mail: edward.sion@villanova.edu}

\author{J. B. Holberg}
\affil{Lunar and Planetary Laboratory,
University of Arizona,
Tucson, AZ 75201,
email: holberg@vega.lpl.arizona.edu}

\author{Terry D. Oswalt}
\affil{Dept. of Physics and Space Sciences,
Florida Institute of Technology,
Melbourne, FL 19085,
email: toswalt@fit.edu}

\author{George P. McCook}
\affil{Dept. of Astronomy and Astrophysics,
Villanova University,
Villanova, PA 19085,
email: george.mccook@villanova.edu}

\author{Richard Wasatonic}
\affil{Dept. of Astronomy and Astrophysics,
Villanova University,
Villanova, PA 19085,
email: richard.wasatonic@villanova.edu}

\begin{abstract}

We present the kinematical properties, distribution of spectroscopic
subtypes, stellar population subcomponents of the white dwarfs within 20
pc of the sun. We find no convincing evidence of halo white dwarfs in the
total 20 pc sample of 129 white dwarfs nor is there convincing evidence of
genuine thick disk subcomponent members within 20 parsecs. Virtually the
entire 20 pc sample likely belongs to the thin disk. The total DA to
non-DA ratio of the 20 pc sample is 1.6, a manifestation of deepening envelope
convection which transforms DA stars with sufficiently thin H surface
layers into non-DAs. The addition of 5 new stars to the 20 pc sample
yields a revised local space density of white dwarfs of $4.9\pm0.5 \times
10^{-3}$/pc$^{3}$ and a corresponding mass density of 
$3.3\pm0.3 \times 10^{-3}$ M$_{\sun}$/pc$^{3}$. We find that at least 15\% of the
white dwarfs within 20 parsecs of the sun (the DAZ and DZ stars) have
photospheric metals that possibly originate from accretion of circumstellar
material (debris disks) around them. If this interpretation is correct, this
suggests the possibility that the same percentage have planets or asteroid-like bodies orbiting them.
\end{abstract}

Subject Headings: techniques: photometric - stars: techniques:
spectroscopic - stars: white dwarfs - stars: statistics - stars:
kinematics

\section{Introduction}

The population of local white dwarfs is astrophysically important for a
number of reasons. First, from complete samples, it offers an excellent
probe of the coolest, least luminous (oldest) component of the overall
white dwarf population. Second, it samples the mix of stellar populations
that evolve into the different spectroscopic subtypes in the immediate
vicinity of the sun.  Third, it provides a unique way of measuring the
local space density and mass density of white dwarfs which are currently
of critical interest because: (a) they represent a history of star
formation and stellar evolution in the Galactic plane; (b) the luminosity
function of these stars can be used to place a lower limit on the age of
the Galactic disk (Liebert, Dahn, Monet 1988, Oswalt et al. 1996); (c)  
cool white dwarfs have been suggested as the origin of the MACHO lensing
objects seen in lensing surveys (Oppenheimer et al. 2001; Kawaler 1996,
Graet et al. 1998); and (d)  they are important to understanding the
overall mass density of the Galactic plane (Bahcall 1984).
   
Recently, Holberg, Sion, Oswalt, McCook, Foran, \& Subsavage (2008,
hereafter, LS08) completed a detailed survey of the local population of
white dwarfs lying within 20 pc of the sun which they estimated to be 80\%
complete. Their sample contained 124 individual degenerate stars,
including both members of four unresolved double degenerate binaries, one
of which was a suspected new double degenerate binary (WD 0423+120).  
Since the publication of LS08, we have added seven additional white dwarfs
to the 20 pc sample, bringing the present total to 131 degenerate stars.
Two of these additional stars are close Sirius-Like companions to nearby K
stars (Holberg 2008), GJ86 (WD0208-510) and HD27442 (WD0415-594),
discovered during exoplanet investigations (see Mugrauer \& Neuhäuser 2005
and Chauvin et al. 2007 respectively).  Of the remaining five single white
dwarfs, three stars, WD0011-721, WD0708-670 and WD1116-470, are from
Subasavage et al. (2008), and one star, WD1315-781, is
from Subasavage et al. (2009).  We have excluded one interesting high
velocity star, WD1339-340.  Lepine, Rich \& Shara (2005) noted that this
star had a space motion of 313 km/s$^{-1}$ relative to the sun based on an
assumed distance of 18 pc.  Holberg et al. (2008) determined that
WD1339-340 has an estimated distance of $21.2\pm 3.5$ pc placing it
formally outside the limits of our 20 pc sample.  Nevertheless, this star
has a finite probability of being within 20 pc.

Including the new stars in the local sample of 129 white dwarfs within 20
pc, yields a revised white dwarf space density of $4.9 \pm 0.5\times
10^{-3}$ pc$^{-3}$. The corresponding mass density is $3.3 \pm
0.3\times10^{-3}$ M$_{\sun}$/pc$^3$.  The completeness of the sample,
however, remains at 80\% since the addition of the companion of GJ86 (at
10.8 pc) contributes to the number of white dwarfs within 13 pc upon which
the stellar density is based.

In this work, we use the enlarged local sample to examine the kinematical
properties, distribution of spectroscopic subtypes, and stellar population
subcomponents of the white dwarfs sampled in this volume of space around
the sun.
 
\section{Distribution of Spectroscopic Subtypes in the Local White Dwarf
Population}

Table 1 presents the sample of white dwarfs within 20 pc of the sun. The
basic observational data from which space motions have been computed are
given in Table 1, which contains by column: (1) the WD-number, (2) the
coordinates (RA and DEC are in decimal degrees), (3) DIST (distance in
parsecs), (4) PM (proper motion in arcsec/yr), (5) PA (position angle in
degrees), and (6) the method of distance determination, denoted by $p$ for
trigonometric parallax, $s$ for spectrophotometric distances and $a$ for a
weighted average of the parallax and photometric distances according to their
respective uncertainties. Distance estimates are taken from LS08 which are based
on both trigonometric parallaxes and spectrophotometric distances.
Whenever possible, preference was given to trigonometric parallax
distances. In LS08 the photometric distances were
computed based upon spectroscopic and photometric measurements using the
techniques described in Holberg, Bergeron \& Gianninas (2008).

Proper Motions are taken from the McCook and Sion Catalog, or where
available, were determined from NOMAD. Radial velocities were available
from the literature for approximately 50\% of our sample, and correspond
either to direct measurements of the white dwarf or the system velocities
or radial velocities of the main sequence companions.  For radial
velocities derived from individual white dwarfs, corrections for the
gravitational redshift have been made based on the individual masses and
radii of each star.  These mass and radius determinations were
interpolated from within the synthetic photometric tables described in
Holberg \& Bergeron (2006) and were based on temperatures and gravities
given in LS08.

Using the spectral types given in LS08 and revised by Holberg (2008), we
summarize the percentage breakdown of spectral subtypes among the 20 pc
sample of local white dwarfs in Table 2.  As expected, the DA stars
dominate the sample with the other spectral groups having roughly the same
percentages as the proper-motion-selected white dwarf sample as a whole.

The DAZ stars which exhibit photospheric metal lines due to accreted
metals are counted separately from the "pure" DA stars in the 20 pc
sample. The 11 DAZ stars within 20 pc account for 8\% of the local sample
while the DZ stars, which have helium-rich atmospheres account for 7\% of
the 20 pc sample. There is only one DB star within 20 parsecs, the cool
DBQZ star LDS678A while the DQ stars, which exhibit molecular and atomic
carbon, account for 9\%.

\begin{deluxetable}{lcr}
\tablecaption{Distribution of WD Spectral Subtypes within 20 Parsecs}
\tablewidth{0pc}
\tablenum{2}
\tablehead{
\colhead{Spectral Type}&
\colhead{Number of Stars}&
\colhead{\% of total}
 }
\startdata
DP,DH  &      17    &    13\% \\
DA     &      58    &    45\%\\
DAZ     &     11     &    8\% \\
DZ       &     9      &   7\%   \\    
DQ        &   12       & 9\%\\
DC         &  21        &16\%\\
DBQZ        &  1        & 1\%\\
\enddata
\end{deluxetable}

The value of the total DA to non-DA ratio of the 20 pc sample, obtained here, 
is 1.6. This ratio holds
important astrophysical significance since the competition between
accretion, diffusion and convective mixing controls and/or modifies the
flow of elements in a high gravity atmosphere and hence determine what
elements are spectroscopically detected at the surface 
(Strittmatter \& Wickramasinghe 1971). The value of the
ratio obtained here is expected for a sample dominated by cool degenerates since earlier
observational studies showed that there is a gradual reduction of this
ratio from 4:1 for white dwarfs hotter than 20,000K down to roughly 1:1
for the coolest degenerates below 10,000K (Sion 1984; Greenstein 1986).  
This transformation in the DA/non-DA ratio occurs when mixing by deepening
envelope convection increasingly transforms DA white dwarfs with
sufficiently thin surface H layers into non-DA stars (Sion 1984;
Greenstein 1986 and references therein). However, DA stars begin their
cooling evolution with a range of hydrogen layer masses. Those with thick
hydrogen layers may not transform via convective mixing and dilution and
will remain DA until the Balmer lines fade below 5000K while DA stars with
sufficiently thin hydrogen layers should undergo transformation to non-DA
stars as the deepening H convection reaches the deeper, more massive
underlying helium convection zone. Thus, some cooling DA stars may
transform to DZ stars when the hydrogen is mixed downward and diluted by
the deepening helium convection. Such an object would be classified DZA
and may owe their observed H abundances to convective mixing and dilution
instead of accretion

\section{Kinematics of the Local White Dwarfs}

For the local white dwarfs with sufficient kinematical information
(photometric or trigonometric parallax, proper motion, position angle) we
have computed the vector components of the space motion {\it U, V} and
{\it W} relative to the Sun in a right-handed system following Wooley et
al. (1970) where {\it U}\ is measured positive in the direction of the
galactic anti-center, {\it V}\ is measured positive in the direction of
the galactic rotation and {\it W}\ is measured positive in the direction
of the north galactic pole.

The space motions were calculated in two ways: (1) using any known radial
velocities for 55 of the 129 stars in the 20 pc sample and; (2) by
assuming zero radial velocity for the entire sample. Also calculated were
the average velocity, velocity dispersions and standard errors for the
entire sample of 129 WDs and for each spectroscopic subclass separately.
We found relatively little kinematical difference among the samples if we
used the 55 available radial velocities and the sample with the assumption
of zero radial velocity. This finding is consistent with earlier studies
(Silvestri et al. 2002; Pauli et al. 2003, 2006) which reported little
difference in the kinematical results with or without the inclusion of
radial velocities.

Table 3 lists by column: the WD-number, the white dwarf spectral type,
$U$-component, $V$-component, $W$-component and $T$, the total space
velocity. All velocities are expressed in km/s. At the end of the
tabulation of individual white dwarf motions, we have listed the average
velocity in each of the three vector components and in the total motion,
T, for the white dwarfs within 20 pc, the velocity dispersion in each
average velocity component and the standard error in each velocity
component.

In order to try to assign stellar population membership, it is useful to
compare the distribution of space velocities of the 20 parsec sample with
other analyses in which the assignment of population membership is on a
secure footing.  Since the population membership of main sequence stars,
unlike white dwarfs, is assigned with chemical abundance data as well as
kinematical characteristics, it is illuminating to compare the
distribution of white dwarfs in the UV velocity plane for the 20 pc sample
with the velocity distribution (velocity ellipses) of a well-studied
sample of main sequence stars. In Fig. 1, we have displayed the $U$ versus
$V$ space velocity diagram for the 20 pc sample of white dwarfs with the
assumption of zero radial velocity relative to three velocity ellipses for
main sequence stars (Chiba \& Beers 2000; see also Vennes \& Kawka 2006).
In the diagram, we have displayed the $2 \sigma$ velocity ellipse contour
(solid line) of the thin disk component, the $2 \sigma$ ellipse of the
thick disk component (short-dashed line) and the $1 \sigma$ contour of the
halo component (long-dashed line).

Examining Table 3, if we take $T > 150$ km/s as the lower cutoff for halo
space motions, then at first glance, six stars or 4\% of the total sample
could be considered likely candidates for the halo population
subcomponent. However, the assignment of white dwarfs to the halo
population subcomponent cannot be made on the basis of space motions
alone. The candidate white dwarfs must also have total stellar ages that
are of order 12 billion years or older. 
In Table 4, we have tabulated the cooling ages of the white dwarfs
in the 20 pc sample in descending order of decreasing age. The cooling
ages listed in Table 4
have been interpolated from the photometric grid
of P.Bergeron (http://astro.umontreal.ca/$\sim$bergeron/CoolingModels/).
Hence, when we take into account the total stellar ages of each of the six stars 
in Table 3 with $T > 150$
km/s, then we conclude that based upon their space motions together with
their total stellar ages, no clear evidence of halo white dwarfs has been
found among the white dwarfs within 20 pc of the sun.

If we take the range of total velocities $60 < T < 150$ km/s to be the
range of space motions of the thick disk subcomponent and if we take total
motions less than 60 km/s to be characteristic of the thin disk, then, at
first glance, 28 white dwarfs (21\% of the 20 pc sample) in Table 1 could
potentially belong to the thick disk with the vast majority of stars
(79\%) belonging to the thin disk.  However, distinguishing thin disk from
thick disk members is a daunting task which involves far more than simply
using space motions alone. Napiwotzki (2008) has already noted that it is
not possible to uniquely identify thick disk stars from thin disk stars
based on space motions in the $U-V$ plane alone or even with all three
vector components available. This is because the space motions of these
thin and thick disk populations overlap. Napiwotzki (2006) used Monte
Carlo simulations of the two populations to estimate the relative
contribution of the two populations to his observed sample. Thus, in the
range 60 km/s $ < v \tan < 150$ km/s, it is not possible to uniquely identify
bona fide thick disk stars without considering their total stellar ages as
well as applying a galactic model as was done by Pauli et al.(2006). Star
formation in the thick disk ended more than 10 billion years ago. It is
clear that white dwarfs which are the descendants of the thick disk
population should have large velocity dispersions similar to those given
by Chiba and Beers (2000) for the thick disk while having lower than
average masses since they are the descendants of low mass progenitors
(Napiwotzki 2009). The velocity dispersions, $\sigma (U)$, $\sigma (V)$,
$\sigma (W)$ of the entire 20 pc sample lies below the boundary line for
thick disk membership which is defined by Chiba and Beers (2000) to be
$\sigma (U) = 46$ km/s, $\sigma (V) = 50$ km/s and $\sigma (W) = 35$ km/s.

This conclusion about the absence of genuine thick disk members in the 20
pc sample is supported by Silvestri et al.'s (2002)  study of 116
common proper motion binaries with white dwarf plus M dwarf components.
Their wide binary pairs gave them at least 3 independent parameters
related to stellar population subcomponent membership: (1) abundance; (2)
intrinsic radial velocity; (3) age (from chromospheric activity, main
sequence-fitting, etc.).  They present kinematics plots (see their figs. 4
and 5) of their wide binary pairs which indicate by comparison with our
Fig. 1 that there are few, if any, halo or thick disk objects in the local
20 pc WD sample. Our local sample has a far smaller velocity dispersion
than the Silvestri et al.(2002) sample.
 
Even in much larger samples of white dwarfs such as the Pauli et al.(2003,
2006) SN Ia Progenitor surveY (SPY), there are relatively few 
genuine halo and thick disk candidates. For example, in their sample of
398 white dwarfs which effectively constituted a magnitude-limited sample,
they examined both the $UVW$ space motions and the Galactic orbits of
their stars.  They found that only 2\% of their sample kinematically
belonged to the halo and 7\% to the thick disk.

Other explanations for high velocity white dwarfs exist. For example,
Rappaport et al. (1994) and Davies, King \& Ritter (2001) have shown that
Type II supernovae may disrupt binaries with orbital periods in the range
of 0.3 to 2 days, yielding single stars whose space velocities are similar
to their original presupernova orbital velocities, $V_{orb}$. Thus a
population of high velocity white dwarfs can be expected to arise from
within the thin disk component of the galactic disk, at least from this
mechanism. High velocity stars may also arise from much wider binary pairs
(see Kawka et al. 2006).

In Table 5, we have tabulated the velocity statistics broken down by
spectral subgroup. By column is listed: (1) the spectroscopic subgroup,
(2) $N$, the number of white dwarfs of a given spectral type, (3) the
vector components of velocity $U, V, W$ and the total motion $T$ (4) the
average velocity in each component, (5) the velocity dispersion of each
component and (6) the error in each component. The motions of individual
spectroscopic subgroups, as seen in Table 5, differ little from each other
although compared with the size of the DA sample, the non-DA groups are
relatively small in number.  The only subgroup of white dwarfs differing
significantly at least from the non-DA subgroups are the magnetic white
dwarfs, which have significantly lower space velocities. Here again,
caution is advised because there are only 16 magnetic degenerates with
known space motions within 20 parsecs of the sun. Nonetheless the trend is
in the direction expected from earlier kinematical evidence (Sion et
al. 1988; Anselowitz et al. 1996) that they are the progeny of more
massive stars. We find a smaller fraction of magnetic white dwarfs, 12\%,
of the 20 pc sample than Kawka \& Vennes (2006) who found 20\% magnetics
within 13 pc of the sun. However, given the quoted uncertainty in the
percentage of magnetic degenerates in their sample and the lower
completeness of the 20 pc sample, we cannot attach any significance to the
differences at the present time.

Finally, although the number of stars in each non-DA spectroscopic
subgroup may be too small for a statistically significant comparison, it
appears that the magnetic white dwarfs have significantly lower velocities
and velocity dispersions than the non-DA spectral types. However, magnetic
white dwarfs closely resemble the kinematical properties of the DA white
dwarfs within 20 pc.

\section{Discussion}              

We have attempted to fully characterize the white dwarfs within 20 pc of
the sun since LS08. By adding 5 new objects within 20 pc, there is now a
total of 129 individual degenerate stars. Three of the added objects are
from Subasavage et al. (2008), one object is from Subasavage et al.(2009)
and two are in wide binaries (Holberg 2008). The completeness of the 20 pc
sample remains at 80\%. Based upon the sample of white dwarfs within 13 pc
which is likely 100\% complete, the local space density of white dwarfs has been
revised to $4.9 \pm 0.5 \times 10^{-3}$ pc$^{-3}$.

In this volume-limited sample, every major white dwarf spectral type (DA,
DC, DQ, DZ, DH/DP) is found except for the pure DB and DO stars. There is,
however, one cool, hybrid DB, the DBQZ star LDS678A within 20 pc. Based
upon their individual space motions, velocity averages, velocity
dispersions and total stellar ages, the 20 pc sample of white dwarfs
overwhelmingly consists of members of the thin disk population. There is
no clear evidence of either halo population white dwarfs or bona fide
thick disk members. It also appears that the magnetic white dwarfs have 
significantly lower velocities and lower velocity dispersions than the 
non-DA spectral types but similar velocity dispersions to the DA stars with 20 pc.  

We have taken a census of DAZ and helium-rich DZ stars within 20 pc of the
sun. The DAZ stars which exhibit accreted metals and which have shorter
diffusion timescales than non-DA stars, have been counted separately from
the ''pure'' DA stars in our statistics. However, we note that there are
11 such objects or 8\% of the local white dwarfs within 20 pc. Because of
their short diffusion timescales, they must be accreting presently or in
the very recent past. This fact coupled with the complete lack of any
local interstellar clouds of sufficient density (Aannestad \& Sion 1985;  
Aannestad, Kenyon, Hammond \& Sion 1993) implies they must be accreting
from circumstellar material. This has been confirmed by the detection with
Spitzer Space Telescope of debris disks around several DAZ stars (Farihi
et al. 2009 and references therein). The accretion rates required to
explain their observed metal abundances are as little as $3\times 10^{8}$
g s$^{-1}$.

The helium-rich DZ stars which typically exhibit calcium and magnesium in
the optical spectra account for 7\% of the 20 pc sample. They have much
longer diffusion timescales (and deeper convection zones) than DAs.  As
cooling helium-rich white dwarfs cool below 12,000K, helium lines can no
longer be detected. Since H is the
lightest element, it tends to float up to the upper layers of the
photosphere. Those DZs with detected H, known as DZA stars, may also have
accreted H and metals from the interstellar medium or from circumstellar
debris disks. However, DZ stars contain less hydrogen in their atmospheres
than what is expected if the accreted material had a solar composition.
Thus, the hydrogen accretion rate must be reduced relative to that of
metals in order to account for the relatively low hydrogen abundances
(with respect to heavier elements) observed in DZ stars (Dufour et al.
2007). While the Dupuis et al. (1993) two-phase model of interstellar
accretion onto white dwarfs cannot be ruled out completely, the cooler DZ
stars do not have the strong UV radiation field needed to ionize accreting
H so that a weakly magnetic, slowly rotating DZ would avoid H accretion by
the propellor mechanism.

Despite the detection of debris/dust disks around some but not all metal
polluted white dwarfs observed by Spitzer, it appears highly probable that
virtually all of the DAZ stars owe their metals to circumstellar
accretion. Thus, from our 20 pc census of white dwarfs, if we consider
only the DAZ stars, then we estimate a lower limit percentage of at least
8\% of the white dwarfs within 20 parsecs of the sun that probably have
circumstellar material (debris disks) around them.  Insofar as we can
regard the DAZ white dwarfs as probes of their circumstellar environments,
then this would suggest a high probability that at least the same
percentage should have asteroid-like minor planets and possibly even
terrestrial-like rocky metallic planets. Indeed one local white dwarf
G29-38 (WD2326+049) in our 20 pc sample is known to have a dusty debris
disk (Zuckerman \& Becklin 1987) while three white dwarfs in our 20 pc
sample are in systems with confirmed extrasolar planets, WD0208-510 (K1V +
DA10), WD0415-594 (K2IV + DA3.8 ) and WD1620-391 (G2V + DA2), although the 
planets in the latter three systems do not orbit the white dwarf. Furthermore,
if we suppose that the DZ stars, including the DZA stars (but see Section
2), have also accreted circumstellar material (have debris disks), then
summing the DZ stars with the DAZ stars, we speculate that at least 15\%
of the white dwarfs within 20 pc of the sun have circumstellar debris
disks with rocky, metallic debris.

We have excluded the DQ stars from this estimate. While it is possible
that the DQ white dwarfs may accrete from circumstellar or even
interstellar matter, we do not include them in our estimate because we
regard it more likely that the source of their photospheric carbon is
either due to convective mixing (Dufour et al.2005 and references therein)  
or is primordial (Dufour et al. 2007). Indeed, only one DQ has been found
to have atmospheric hydrogen (G99-37). Accreted metals are also easier
to hide in DQ atmospheres due to the increased opacity provided by carbon.

\acknowledgments

We are grateful to an anonymous referee for numerous helpful suggestions
and comments. This work was supported by NSF grant AST05-07797 to the
University of Arizona, Villanova University and the Florida Institute of
Technology. Participation by TDO was also supported by NSF grant
AST-0807919 to FIT.

\begin{deluxetable}{lccccccc}
\tablecaption{Observational Data Used in Space Motions}
\tablewidth{0pc}
\tablenum{1}
\tablehead{ 
\colhead{WD}&
\colhead{TYPE}&
\colhead{RA}&
\colhead{DEC}&
\colhead{DIST}&
\colhead{PM}&
\colhead{PA}&
\colhead{method}
}
\startdata 
0000-345 & DCP9   & 000.677 & -34.225 & 12.65 & 0.899 & 217.337& a\\
0008+423 & DA6.8 & 002.842 & +42.668 & 17.94 & 0.237 & 193.4 & s \\
0009+501 & DAH7.7 & 002.413 & +50.428 & 11.03 & 0.718 & 216.0 & p \\
0011-134 & DCH8.4 & 003.554 & -13.177 &  19.49  &  0.911 & 217.7 & p \\
0011-721 & DA8.0 &  003.457 & -71.831 & 17.80 & 0.326 & 141.300& s \\
0038-226 & DQ9.3 &  010.354 & -22.347 & 09.88 & 0.567 & 229.004& p \\
0046+051 & DZ8.1 & 012.291 & +05.388 & 04.32 & 2.978 & 155.538& p \\
0108+277 & DAZ9.6 & 017.686 & +27.970 & 13.79 & 0.227 & 219.321& s  \\
0115+159 & DQ6 & 019.500 & +16.172 & 15.41 & 0.648 & 181.805 & p \\
0121-429 & DAH7.9 & 021.016 & -42.773 & 17.67 & 0.540 & 155.143 & p \\
0135-052 & DA6.9 & 024.497 & -04.995 & 12.35 & 0.681 & 120.838 & p \\
0141-675 & DA7.8 & 025.750 & -67.282 & 09.70 & 1.048 & 198.279 & s \\
0148+467 & DA3.8 & 028.012 & +47.001 & 16.06 & 0.124 & 000.568 & a \\
0148+641 & DA5.6 & 027.966 & +64.431 & 17.13 & 0.285 & 123.857 & s \\
0208+396 & DAZ7.0 & 032.836 & +39.922 & 16.13 & 1.145 & 115.746 & a \\
0208-510 & DA10 & 032.500  & -50.133 &  10.8 & 2.192 &  72.666 & p \\ 
0213+427 & DA9.4 & 034.281 & +42.977 & 19.67 & 1.047 & 125.065 & a \\
0230-144 & DC9.5 & 038.157 & -14.197 & 15.38 & 0.687 & 177.114 &  a \\
0233-242 & DC9.3 & 038.840 & -24.013 & 15.67 & 0.622 & 189.015 & s \\
0245+541 & DAZ9.7 & 042.151 & +54.383 & 10.35 & 0.573 & 227.827 & p  \\
0310-688 & DA3.1 & 047.628 & -68.600 & 10.15 & 0.111 & 158.097 & p \\
0322-019 & DAZ9.7 & 051.296 & -01.820 & 16.81 & 0.909 & 164.625 & p \\
0326-273 & DA5.4 & 052.203 & -27.317 & 19.73 & 0.850 & 071.629 & s \\
0341+182 & DQ7.7 & 056.145 & +18.436 & 19.01 & 1.199 & 150.771 & p \\
0344+014 & DQ9.9 & 056.778 & +01.646 & 19.90 & 0.473 & 150.400 & s \\
0357+081 & DC9.2 & 060.111 & +08.235 & 17.46 & 0.535 & 222.273 & a \\
0413-077 & DAP3.1 & 063.839 & -07.656 & 05.04 & 4.088 & 213.216 & p \\
0415-594 & DA3.8 & 064.122 & -59.302 & 18.23 & 0.174 & 195.838 & p \\
0423+120 & DA8.2 &  066.473 & +12.196 & 17.36 & 0.244 & 335.866 & p \\
0426+588 & DC7.1 & 067.802 & +58.978 & 05.53 & 2.426 & 147.602 & p \\
0433+270 & DA9.3 & 069.187 & +27.164 & 17.85 & 0.276 & 124.196 & p \\
0435-088 & DQ8.0 & 069.447 & -08.819 & 09.51 & 1.574 & 171.103 & p \\
0457-004 & DA4.7 & 074.930 & -00.377 & 17.67 & 0.293 & 142.872 & s \\
0548-001 & DQP8.3 &  087.831 & -00.172 & 11.07 & 0.251 & 025.810 & p \\
0552-041 & DZ11.8 & 088.789 & -04.168 & 06.45 & 2.376 & 166.966 & p \\
0553+053 & DAP8.9 & 089.106 & +05.536 & 07.99 & 1.027 & 204.993 & p \\
0642-166 & DA2 & 101.288 & -16.713 & 02.63 & 1.339 & 204.057 &  p \\
0644+025 & DA8 & 101.789 & +02.517 & 17.83 & 0.423 & 272.571 & a \\
0644+375 & DA2.5 & 101.842 & +37.526 & 15.41 & 0.962 & 193.561 & p \\
0655-390 & DA8.0 & 104.274 & -39.159 & 17.20 & 0.340 & 242.600 & s \\
0657+320 & DC10.1 & 105.215 & +31.962 & 15.19 & 0.691 & 149.362 & p \\
0659-063 & DA7.8 & 105.478 & -06.463 & 12.13 & 0.898 & 184.980 & a \\
0708-670 & DC & 107.217 & -67.108 & 17.50 & 0.246 & 246.300 & s \\
0727+482.1 & DA10.0 &  112.678 & +48.199 & 11.01 & 1.286 & 190.069 & a \\
0727+482.2 & DA10.0 & 112.697 & +48.173 & 11.20 & 1.286 & 190.069 & a \\
0728+642 & DAP11.2 & 113.378 & +64.157 & 13.40 & 0.266 & 171.352 & s \\
0736+053 & DQZ6.5 & 114.827 & +05.227 & 03.50 & 1.259 & 214.574 & p \\
0738-172 & DAZ6.7 & 115.086 & -17.413 & 09.28 & 1.267 & 115.137 & p \\
0743-336 & DC10.6 & 116.396 & -34.176 & 15.20 & 1.736 & 352.670 & p  \\
0747+073.1 & DC12.1 &  117.563 & +07.193 & 18.25 & 1.804 & 173.414 & a \\
0747+073.2 & DC11.9 &  117.563 & +07.193 & 18.25 & 1.804 & 173.414 & a \\
0749+426 & DC11.7 & 118.305 & +42.500 & 19.74 & 0.420 & 165.845 & s \\
0751-252 & DA10.0 & 118.485 & -25.400 & 18.17 & 0.426 & 300.200 & p \\
0752-676 & DC10.3 & 118.284 & -67.792 & 07.05 & 2.149 & 135.866 & a \\
0806-661 & DQ4.2 & 121.723 & -66.304 & 19.17 & 0.398 & 132.700 & p \\
0821-669 & DA9.8 &  125.361 & -67.055 & 10.65 & 0.758 & 327.600 & p \\
0839-327 & DA5.3 & 130.385 & -32.943 & 08.07 & 1.600 & 322.056 & a \\
0840-136 & DZ10.3 & 130.701 & -13.786 & 19.30 & 0.272 & 263.000 & s \\
0912+536 & DCP7 & 138.983 & +53.423 & 10.31 & 1.563 & 223.997 & p \\
0955+247 & DA5.8 & 149.451 & +24.548 & 18.83 & 0.420 & 219.848 & s \\
1009-184 & DZ7.8 & 153.007 & -18.725 & 18.00 & 0.519 & 268.200 & p \\
1019+637 & DA7.3 & 155.787 & +63.461 & 13.93 & 0.379 & 053.160 &  s \\
1033+714 & DC9 & 159.260 & +71.182 & 20.00 & 1.917 & 256.008 & s \\
1036-204 & DQP10.2 & 159.731 & -20.682 & 14.29 & 0.628 & 333.300 &  p \\
1043-188 & DQ8.1 & 161.412 & -19.114 & 17.57 & 1.978 & 251.636 & a \\
1055-072 & DA6.8 & 164.396 & -07.523 & 11.96 & 0.827 & 276.328 & a \\
1116-470 & DC  & 169.613 & -47.365 & 17.90 & 0.322 & 275.100 & s \\
1121+216 & DA6.7 &  171.054 & +21.359 & 13.55 & 1.040 & 269.240 & a \\
1124+595 & DA4.8 & 171.171 & +59.321 & 17.90 & 0.156 & 108.203 & s \\
1132-325 & DC & 173.623 & -32.832 & 09.54 & 0.940 & 038.954 & p \\
1134+300 & DA2.5 & 174.271 & +29.799 & 15.37 & 0.148 & 267.948 & a \\
1142-645 & DQ6.4 & 176.428 & -64.841 & 04.62 & 2.687 & 097.414 & p \\
1202-232 & DAZ5.8 & 181.361 & -23.553 & 10.82 & 0.229 & 009.068 & p \\
1223-659 & DA6.5 & 186.625 & -66.205 & 16.25 & 0.153 & 195.124 & p \\
1236-495 & DA4.4 & 189.708 & -49.800 & 13.71 & 0.490 & 255.708 & a \\
1257+037 & DA8.7 & 195.037 & +03.478 & 16.18 & 0.969 & 206.195 & a \\
1309+853 & DAP9 & 197.171 & +85.041 & 18.05 & 0.321 & 140.811 & p \\
1310-472 & DC11.9 & 198.248 & -47.468 & 14.95 & 2.204 & 105.252 & a \\
1315-781 & DA & 119.857 & -78.239 & 19.23 & 0.470 & 139.5   & p \\
1327-083 & DA3.7 & 202.556 & -08.574 & 16.47 & 1.204 & 246.761& a \\
1334+039 & DZ10.0 & 204.132 & +03.679 & 08.24 & 3.880 & 252.774 & p \\
1344+106 & DAZ7.1 & 206.851 & +10.360 & 19.50 & 0.903 & 260.569 & a \\
1345+238 & DA10.7 & 207.125 & +23.579 & 12.06 & 1.496 & 274.636 &  p \\
1444-174 & DC10.1 & 221.855 & -17.704 & 14.07 & 1.144 & 252.643 & a \\
1544-377 & DA4.7 & 236.875 & -37.918 & 13.16 & 0.468 & 242.838 & s \\
1609+135 & DA5.8 & 242.856 & +13.371 & 18.35 & 0.551 & 178.513 & p \\
1620-391 & DA2 & 245.890 & -39.918 & 12.87 & 0.075 & 089.962 & p \\
1626+368 & DZ5.5 & 247.104 & +36.771 & 15.95 & 0.888 & 326.667 & p \\
1632+177 & DA5 & 248.674 & +17.609 & 15.99 & 0.088 & 108.434 & s \\
1633+433 & DAZ7.7 & 248.755 & +43.293 & 15.11 & 0.373 & 144.151 & p \\
1633+572 & DQ8.2 & 248.589 & +57.169 & 14.45 & 1.644 & 317.229 & p \\
1647+591 & DAV4.2 & 252.106 & +59.056 & 10.79 & 0.323 & 154.498 & a \\
1653+385 & DAZ8.8 & 253.690 & +38.493 & 15.35 & 0.328 & 177.596 &  s \\
1655+215 & DA5.4 & 254.291 & +21.446 & 18.63 & 0.582 & 178.040 & a  \\
1705+030 & DZ7.1 & 257.033 & +02.962 & 17.54 & 0.379 & 180.907 & p \\
1748+708 & DQP9.0 & 267.033 & +70.876 & 06.07 & 1.681 & 311.394 & p \\
1756+827 & DA7.1 & 270.359 & +82.745 & 15.55 & 3.589 & 336.541 & a \\
1814+134 & DA9.5 & 274.277 & +13.473 & 14.22 & 1.207 & 201.500 & p \\
1820+609 & DA10.5 & 275.332 & +61.018 & 12.79 & 0.713 & 168.516 & p \\
1829+547 & DQP7.5 & 277.584 & +54.790 & 14.97 & 0.399 & 317.233 &  p \\
1900+705 & DAP4.5 & 285.042 & +70.664 & 12.99 & 0.506 & 010.466 & p \\
1917+386 & DC7.9 & 289.744 & +38.722 & 11.70 & 0.251 & 174.028 &  p \\
1917-077 & DBQA5 & 290.145 & -07.666 & 10.08 & 0.174 & 200.602 & p \\
1919+145 & DA3.5 & 290.417 & +14.673 & 19.80 & 0.074 & 203.805 & p \\
1935+276 & DA4.5 & 294.307 & +27.721 & 18.00 & 0.436 & 088.686 & a \\
1953-011 & DAP6.5 & 299.121 & -01.042 & 11.39 & 0.827 & 212.314 & p \\
2002-110 & DA10.5 & 301.395 & -10.948 & 17.33 & 1.074 & 095.523 & p \\
2007-303 & DA3.3 & 302.736 & -30.218 & 15.37 & 0.428 & 233.492 & p \\
2008-600 & DC9.9 & 303.132 & -59.947 & 16.55 & 1.440 & 165.500 & p \\
2032+248 & DA2.5 & 308.591 & +25.063 & 15.65 & 0.692 & 215.554 & a \\
2047+372 & DA4 & 312.277 & +37.470 & 17.77 & 0.219 & 047.150 & s \\
2048+263 & DA9.7 & 312.586 & +26.511 & 19.89 & 0.514 & 235.044 &  a \\
2054-050 & DC10.9 & 314.199 & -04.844 & 17.06 & 0.802 & 106.562 &  p \\
2105-820 & DAP4.9 & 318.320 & -81.820 & 18.12 & 0.516 & 146.371 & a \\
2117+539 & DA3.5 & 319.734 & +54.211 & 17.88 & 0.213 & 336.371 & a \\
2138-332 & DZ7 & 325.489 & -33.008 & 15.63 & 0.210 & 228.500 &   \\
2140+207 & DQ6.1 & 325.670 & +20.999 & 12.52 & 0.681 & 199.444 & p \\
2154-512 & DQ7 & 329.410 & -51.008 & 16.36 & 0.374 & 184.738 & p \\
2159-754 & DA5 & 331.087 & -75.223 & 14.24 & 0.529 & 275.635 & s \\
2211-392 & DA8 & 333.644 & -38.985 & 18.80 & 1.056 & 110.100 & p \\
2226-754 & DC9.9 & 337.662 & -75.232 & 15.11 & 1.868 & 167.500 & s \\
2226-755 & DC12.1 & 337.639 & -75.256 & 15.11 & 1.868 & 167.500 & s \\
2246+223 & DA4.7 & 342.273 & +22.608 & 19.05 & 0.525 & 083.551 & p \\
2251-070 & DZ13 & 343.472 & -06.781 & 08.08 & 2.585 & 105.369 & p \\
2322+137 & DA10.7 & 351.332 & +14.060 & 18.76 & 0.037 & 071.565 & s \\
2326+049 & DAZ4.4 & 352.198 & +05.248 & 13.62 & 0.493 & 236.406 & p \\
2336-079 & DAZ4.6 & 354.711 & -07.688 & 15.94 & 0.192 & 172.208 & p \\
2341+322 & DA4.0 & 355.961 & +32.546 & 18.33 & 0.229 & 252.150 & a \\
2359-434 & DAP5.8 & 000.544 & -43.165 & 07.27 & 1.020 & 135.198 & p 
\enddata
\tablerefs{Comments
\\ s - spectrophotometric
\\ p - trigonometric parallax
\\ a - weighted mean average}
\end{deluxetable}

\begin{deluxetable}{llrrrr}
\tablecaption{Space Motions of White Dwarfs Within 20 pc }
\tablewidth{0pc}
\tablenum{3}
\tablehead{
\colhead{WD No.}&
\colhead{Type}&
\colhead{U}&
\colhead{V}&
\colhead{W}&
\colhead{T}
 }
 \startdata
   0000-345 &    DCP9   &   -11.9&    -43.7 &     3.2&     45.4\\
   0008+423  &   DA6.8  &   -10.0 &    -1.9 &   -17.4 &    20.2\\
   0009+501   &  DAH7.7 &   -28.1 &     8.6 &   -24.1  &   38.0\\
   0011-134   &  DCH8.4 &     -75.5 &   -31.1&     -9.8  &   82.3\\
   0011-721  &   DA8.0   &   10.3 &   -23.2 &    12.2 &    28.2\\
   0038-226  &   DQ9.3 &    -24.9 &    -4.6 &    -1.2 &    25.3\\
   0046+051   &  DZ8.1  &      -2.8 &   -53.6 &   -30.3 &    61.6\\
   0108+277   &  DAZ9.6  &  -11.9 &     0.3 &    -9.9  &   15.5\\
   0115+159   &  DQ6   &    -21.4 &   -27.2 &   -32.0 &    47.2\\
   0121-429   &  DAH7.9 &    -0.7 &   -46.8  &   11.1 &    48.1\\
   0135-052  &   DA6.9 &     16.8&    -37.2 &    -1.6 &    40.9\\
   0141-675  &   DA7.8   &    -21.8 &   -22.8 &    21.5 &    38.2\\
   0148+467  &   DA3.8  &     2.6  &    1.8  &    8.3 &     8.8\\
   0148+641  &   DA5.6 &     12.3 &   -11.2  &   -6.8  &   18.0\\
   0208+396  &   DAZ7.0  &   43.0 &   -62.9 &    -6.7 &    76.5\\
   0208-510  &   DA10    &   85.6 &   -42.1  &  24.0   &   98.4\\   
   0213+427 &    DA9.4  &    37.8 &   -67.3  &  -19.8  &   79.8\\
   0230-144   &  DC9.5  &   -20.7 &   -43.3 &   -13.3  &   49.9\\
   0233-242   &  DC9.3  &   -24.2 &   -34.1 &    -9.8  &   43.0\\
   0245+541   &  DAZ9.7  &   -16.4  &   10.9 &   -24.2  &   31.3\\
   0310-688   &  DA3.1   &    0.1  &   -4.2  &    2.8 &     5.1\\
   0322-019    & DAZ9.7 &   -21.5 &   -66.8 &   -20.7 &    73.2\\
   0326-273   &  DA5.4 &     46.7 &   -29.7&     47.2 &    72.8\\
   0341+182   &  DQ7.7   &  -16.0 &   -94.9  &  -23.8 &    99.1\\
   0344+014   &  DQ9.9 &     -7.9 &   -43.8  &   -4.9 &    44.8\\
   0357+081   &  DC9.2  &   -25.3  &   -4.3 &   -36.8 &    44.9\\
   0413-077   &  DAP3.1  &     -44.2 &   -34.8 &   -73.0 &    92.2\\
   0415-594  &   DA3.8  &    -4.9&     -5.7&     -1.6  &    7.7\\
   0423+120  &   DA8.2  &       5.3 &    18.1  &    3.8 &    19.3\\
   0426+588   &  DC7.1    &     0.9&    -35.4 &    -3.7 &    35.6\\
   0433+270   &  DA9.3  &    -0.1  &  -19.8  &    7.0  &   21.1\\
   0435-088   &  DQ8.0 &    -25.1 &   -63.2 &   -21.6 &    71.4\\
   0457-004  &   DA4.7  &    -6.1&    -23.8  &    2.5 &    24.7\\
   0548-001   &  DQP8.3  &     4.6  &    6.3 &    10.5 &    13.1\\
   0552-041   &  DZ11.8 &    -25.3 &   -65.7 &   -20.2 &    73.3\\
   0553+053   &  DAP8.9 &   -12.2 &   -19.8  &  -31.2  &   38.9\\
   0642-166   &  DA2  &      -0.6 &   -10.2 &   -14.5  &   17.7\\
   0644+025   &  DA8    &     9.4 &    15.1 &   -30.9 &    35.7\\
   0644+375   &  DA2.5  &   -19.7  &  -39.6 &   -32.7  &   55.1\\
   0655-390  &   DA8.0     &      2.7 &     2.3 &  -28.2 &   28.4\\
   0657+320   & DC10.1 &     -23.3  & -40.0 &   10.4 &   47.5\\
   0659-063   & DA7.8  &  -19.7 &  -37.5  & -29.5  &  51.6\\
   0708-670  & DC &          4.7 &    4.9&   -19.9 &   21.1\\
   0727+482.1 & DA10.0 &  -14.6 &  -40.5 &  -14.7  &  45.5\\
   0727+482.2 & DA10.0 &  -14.9  & -41.1  & -14.9   & 46.3\\
   0728+642  & DAP11.2 &      -4.5  &  -9.2 &    3.9  &  11.0\\
   0736+053  &  DQZ6.5  &   0.6  & -12.0 &  -18.4 &   21.9\\
   0738-172  &  DAZ6.7  & -30.5 &  -29.2  &  30.2 &   51.9\\
   0743-336   & DC10.6   &    49.4 &   67.6  &  61.4 &  103.9\\
   0747+073.1 & DC12.1    &  -76.7&  -126.3  & -49.0 &  155.7\\
   0747+073.2  &DC11.9  &    -76.7  &-126.3  & -49.0  & 155.7\\
   0749+426   &  DC11.7 &   -17.8 &   -29.6  &    5.6  &   35.0\\
   0751-252 & DA10.0 &               19.1&     15.5 &   -13.8 &    28.3\\
   0752-676   &  DC10.3 &   -29.3 &   -15.4   &  13.6  &   35.9\\
   0806-661   &  DQ4.2   &    -17.2  &   -6.5  &    7.3  &   19.8\\
   0821-669   &  DA9.8  &    16.8   &   4.5  &    4.4   &  18.0\\
   0839-327  &   DA5.3   &   38.2 &    26.6 &     4.6 &    46.8\\
   0840-136  &   DZ10.3 &    14.0  &   -0.0  &  -21.0  &   25.2\\
   0912+536   &  DCP7    &    21.0  &  -42.5  &  -24.3  &   53.3\\
   0955+247  &   DA5.8  &     6.1  &  -28.1 &   -17.7  &   33.8\\
   1009-184  &   DZ7.8     &    36.1 &    -8.8 &   -26.4  &   45.6\\
   1019+637&     DA7.3  &   -14.7 &    17.9 &     3.5   &  23.4\\
   1033+714  &   DC9  &     138.8 &   -73.2 &   -58.0  &  167.3\\
   1036-204  &   DQP10.2 &   36.0 &    17.9 &    20.7  &   45.2\\
   1043-188  &   DQ8.1  &   110.7  &  -71.1 &  -107.8 &   170.1\\
   1055-072  &   DA6.8 &     42.3 &   -10.4 &   -16.3 &    46.5\\
   1116-470  &   DC &        24.3  &   -9.2 &   -7.0 &    27.0\\
   1121+216   &  DA6.7  &    57.8&    -25.3 &   -21.5 &    66.7\\
   1124+595   &  DA4.8  &   -12.2  &    1.3   &   6.0 &    13.7\\
   1132-325  &   DC   &     -11.4 &    22.5  &   35.8 &    43.8\\
   1134+300   &  DA2.5   &    9.2  &   -4.6  &   -2.9  &   10.7\\
   1142-645   &  DQ6.4   &  -52.1 &    25.3&      6.6  &   58.3\\
   1202-232   &  DAZ5.8   &    3.0  &    6.1  &    8.6 &    11.0\\
   1223-659   &  DA6.5    &      0.7 &     0.1  &   -8.4  &    8.4\\
   1236-495   &  DA4.4  &    24.0  &  -17.0   &  -8.6  &   30.7\\
   1257+037   &  DA8.7  &    -6.2 &   -69.4  &  -25.6 &    74.2\\
   1309+853   &  DAP9   &   -16.3  &   -1.3   &  15.4   &  22.5\\
   1310-472   &  DC11.9  &   -133.6 &    81.4 &   -51.5 &   164.8\\
   1315-781   &  DA    &    -23.5  &   +26.4  &  -32.3  &  47.8 \\
   1327-083   &  DA3.7  &    49.2 &   -76.6  &   -7.4  &   91.3\\
   1334+039   &  DZ10.0 &    87.5 &  -122.2  &    8.5  &  150.5\\
   1344+106   &  DAZ7.1   &     54.6  &  -62.3  &   14.1 &    84.1\\
   1345+238   &  DA10.7  &   67.0 &   -47.4  &   20.1  &   84.5\\
   1444-174   &  DC10.1  &   33.0 &   -61.4 &    17.6 &    71.9\\
   1544-377   &  DA4.7  &     5.5 &   -22.4  &    7.6 &    24.3\\
   1609+135   &  DA5.8  &   -25.2 &   -35.3 &   -17.4 &    46.8\\
   1620-391   &  DA2     &   -1.3  &    2.8  &   -3.2  &    4.5\\
   1626+368  &   DZ5.5  &    35.1  &   15.5 &    35.6 &    52.3\\
   1632+177   &  DA5   &      -2.7 &     2.3 &    -5.3   &   6.4\\
   1633+433  &   DAZ7.7  &  -12.7 &    -4.2  &  -13.9  &   19.4\\
   1633+572   &  DQ8.2  &      45.4&     -3.0&     54.5 &    71.0\\
   1647+591   &  DAV4.2  &   -6.0  &   -3.0  &   -5.4  &    8.6\\
   1653+385  &   DAZ8.8 &   -10.4  &  -14.8 &    -5.5 &    18.9\\
   1655+215  &   DA5.4  &   -25.9  &  -35.9  &  -18.4 &    47.9\\
   1705+030  &   DZ7.1 &    -16.6 &   -23.1 &   -13.4 &    31.5\\
   1748+708   &  DQP9.0 &      5.8 &   -10.8 &    34.2  &   36.3\\
   1756+827  &   DA7.1  &     8.8 &   -30.3 &   106.4 &   111.0\\
   1814+134  &   DA9.5  &   -47.7   & -72.0 &    -8.6  &   86.8\\
   1820+609  &   DA10.5 &    -9.8 &    -9.0 &   -19.7 &    23.7\\
   1829+547  &   DQP7.5  &    2.8 &    -0.7 &    24.3 &    24.4\\
   1900+705   &  DAP4.5 &     6.7 &     5.5  &    6.0 &    10.5\\
   1917+386    & DC7.9  &    -5.2  &   -5.7 &    -8.6  &   11.6\\
   1917-077   &  DBQA5  &    -5.4 &    -6.4 &    -0.5 &     8.4\\
   1919+145   &  DA3.5  &    -4.2 &    -5.0 &    -0.7 &     6.6\\
   1935+276   &  DA4.5  &    16.0  &   10.2&    -31.7 &    37.0\\
   1953-011  &   DAP6.5  &     -32.3 &   -31.9  &    3.7 &    45.6\\
   2002-110 &    DA10.5  &   40.4   &   9.8  &  -76.6  &   87.2\\
   2007-303   &  DA3.3 &    -22.0  &  -18.3  &   18.0  &   33.8\\
   2008-600   &  DC9.9 &    -17.7  &  -62.3  &   -3.3  &   64.9\\
   2032+248   &  DA2.5  &   -36.9 &   -24.9 &    -2.8  &   44.7\\
   2047+372   &  DA4  &      13.4 &     5.5 &    -1.4 &    14.6\\
   2048+263   &  DA9.7  &   -37.6 &   -16.2  &   11.8 &    42.7\\
   2054-050  &   DC10.9  &   31.1  &  -11.4 &   -54.1 &    63.5\\
   2105-820  &   DAP4.9  &    12.5 &   -17.4 &     3.6  &   21.8\\
   2117+539   &  DA3.5  &    -0.8  &    2.1 &    18.3 &    18.5\\
   2138-332   &  DZ7  &     -15.0 &    -8.3  &    8.4  &   19.1\\
   2140+207   &  DQ6.1  &     -28.0 &   -18.9 &    -17.1 &    37.9\\
   2154-512   &  DQ7   &    -12.4 &   -22.4  &    9.9  &   27.5\\
   2159-754   &  DA5   &    -27.6 &     7.6  &   18.5 &    34.1\\
   2211-392   &  DA8   &     57.8  &  -43.7  &  -44.5  &   85.1\\
   2226-754   &  DC9.9  &    -0.5 &   -88.5  &   71.5  &  113.8\\
   2226-755   &  DC12.1  &   -0.5 &   -88.5  &   71.5  &  113.8\\
   2246+223   &  DA4.7  &    42.4  &  -10.4  &  -16.9  &   46.8\\
   2251-070   &  DZ13 &      68.0 &   -48.2 &   -50.0 &    97.2\\
   2322+137    & DA10.7   &   3.2   &  -0.5  &   -0.3  &    3.2\\
   2326+049   &  DAZ4.4  &  -31.6   &  -2.1  &   -1.3   &  31.7\\
   2336-079    & DAZ4.6   &  -5.8  &  -13.4 &    -5.9 &    15.8\\
   2341+322   &  DA4.0 &    -18.8  &    5.6   &  -0.4   &  19.6\\
   2359-434   &  DAP5.8  &    13.9  &  -16.4 &   -27.6  &   35.0\\
   &&&&&\\
        &      \bf{U}  &  \bf{V}    &  \bf{W}    & \bf{T}&\\
   AVG. &               +0.9 &     -21.5 &     -5.2 &     47.6\\
   DISP.        &     35.3&       33.1 &     28.1  &    36.6\\
   ERROR          &     3.1    &    2.9    &   2.4   &    3.2
\enddata
\end{deluxetable}

\begin{deluxetable}{llrr}
\tablecaption{Cooling Ages of White Dwarfs Within 20 pc}
\tablewidth{0pc}
\tablenum{4}
\tablehead{
\colhead{WD No.}&
\colhead{T$_{eff}$}&
\colhead{Log g}&
\colhead{Age}
} 
\startdata
WD2251-070&	4000&	8.01	&9.04E+09\\
WD1310-472&	4220&	8.12&	8.83E+09\\
WD2226-755	& 4177	&8&	8.54E+09\\
WD2226-754&	4230	&8	&8.40E+09\\
WD0749+426&	4300	&8&	8.19E+09\\
WD2002-110&	4800&	8.31&	8.14E+09\\
WD1444-174&	4960&	8.37&	8.03E+09\\
WD2054-050&	4620&	8.09&	7.58E+09\\
WD0728+642&	4500&	8&	7.58E+09\\
WD0108+277	& 5270&	8.36&	6.93E+09\\
WD0208-510&	4700&	8&	6.90E+09\\
WD0747+073.1	& 4850	&8.04&	6.53E+09\\
WD0743-336&	4740&	7.97&	6.52E+09\\
WD0552-041&	4270&	7.8&	6.42E+09\\
WD0747+073.2	& 5000&	8.12&	6.41E+09\\
WD0245+541&	5280&	8.28&	6.30E+09\\
WD1033+714	&4888&	8&	6.15E+09\\
WD0657+320&	4990&	8.07&	6.15E+09\\
WD0727+482.2	&5060	&8.12&	6.13E+09\\
WD0840-136&	4900&	8	&6.10E+09\\
WD1036-204&    4948&   8&       5.90E+09\\
WD1748+708&	5590&	8.36&	5.69E+09\\
WD2008-600&	5078&	8	&5.28E+09\\
WD0344+014&	5084&	8	&5.25E+09\\
WD1820+609&	4780&	7.83&	5.24E+09\\
WD1345+238&	4590&	7.76&	5.20E+09\\
WD1334+039&	5030	&7.95&	5.19E+09\\
WD0708-670&	5108	&8	&5.12E+09\\
WD0727+482.1	&5020&	7.92&	5.03E+09\\
WD0751-252&	5159&	8&	4.86E+09\\
WD0821-669&	5160&	8&	4.86E+09\\
WD1829+547&	6280&	8.5	&4.78E+09\\
WD1653+385	&5700&	8.28	&4.77E+09\\
WD0659-063&	6520&	8.71&	4.40E+09\\
WD0752-676&	5730&	8.21&	4.19E+09\\
WD1257+037&	5595&	8.16	&4.18E+09\\
WD0230-144&	5480&	8.11&	4.12E+09\\
WD1814+134&	5313&	8	&4.09E+09\\
WD0553+053&	5790&	8.2&	3.97E+09\\
WD0433+270&	5620&	8.14&	3.97E+09\\
WD0213+427&     5600&    8.12&  3.87E+09\\
WD0000-345&	6240&	8.31&	3.78E+09\\
WD0233-242&	5400&	8&	3.66E+09\\
WD0009+501&	6610&	8.36&	3.58E+09\\
WD0644+025&	7410&	8.66&	3.48E+09\\
WD0011-134&	6010&	8.2&	3.43E+09\\
WD1917+386&	6390&	8.28&	3.42E+09\\
WD0357+081&	5490&	8.02&	3.37E+09\\
WD0038-226&	5400	&7.91&	3.29E+09\\
WD0548-001	&6070	&8.18&	3.24E+09\\
WD0046+051&	6220&	8.19&	3.13E+09\\
WD1055-072&	7420&	8.42&	3.02E+09\\
WD1309+853&	5600&	8&	2.98E+09\\
WD0008+423&	7380	&8.38&	2.89E+09\\
WD1705+030&	6580&	8.2&	2.80E+09\\
WD2159-754&	9040&	8.95&	2.76E+09\\
WD1019+637&	6981  &8.253&	2.69E+09\\
WD0912+536&	7160&	8.28&	2.66E+09\\
WD1633+572&	6180&	8.09&	2.60E+09\\
WD1043-188&	6190&	8.09&	2.59E+09\\
WD1116-470&     5856&    8&     2.50E+09\\
WD2359-434&	8570&	8.6&	2.48E+09\\
WD0426+588&	7120&	8.17&	2.21E+09\\
WD0423+120&	6150&	8&	2.11E+09\\
WD1121+216&	7471&	8.197&	2.08E+09\\
WD1609+135&	9321&	8.644&	2.08E+09\\
WD0141-675&	6460&	8.04&	2.07E+09\\
WD0457-004&	10800&	9.15&	2.02E+09\\
WD2211-392&	6290&	8&	2.00E+09\\
WD1344+106&	7135&	8.119&	1.97E+09\\
WD0121-429&	6369	&8	&1.93E+09\\
WD1953-011&	7920&	8.23&	1.91E+09\\
WD0655-390&	6415&	8	&1.90E+09\\
WD0011-721&	6439&	8	&1.88E+09\\
WD1009-184&	6449&	8&	1.87E+09\\
WD0435-088&	6300&	7.93&	1.87E+09\\
WD0341+182&	6510&	7.99&	1.81E+09\\
WD2246+223&	10647&	8.803&	1.79E+09\\
WD0322-019&	5220&	7.5&	1.79E+09\\
WD0208+396&	7340&	8.1&	1.77E+09\\
WD0955+247&	8621&	8.301&	1.73E+09\\
WD2322+137&	4700	&7&	1.73E+09\\
WD0148+641&	8938	&8.354&	1.70E+09\\
WD1223-659	&7740&	8.13&	1.66E+09\\
WD2048+263&	5200&	7.31&	1.58E+09\\
WD0738-172&	7590	&8.07&	1.51E+09\\
WD1236-495&	11748&	8.802&	1.46E+09\\
WD2138-332&	7188	&8	&1.43E+09\\
WD1633+433	&6518	&7.735&	1.39E+09\\
WD1142-645	&7900&	8.07&	1.36E+09\\
WD1756+827	&7270&	7.98&	1.36E+09\\
WD0115+159	&9050&	8.19&	1.24E+09\\
WD0135-052	&7280	&7.85	&1.20E+09\\
WD0736+053&	7740&	8	&1.18E+09\\
WD1655+215	&9313	&8.203&	1.17E+09\\
WD1202-232	&8774&	8.1	&1.11E+09\\
WD1626+368	&8440&	8.02	&9.95E+08\\
WD1900+705	&12070&	8.58	&9.48E+08\\
WD2140+207&	8200&	7.84	&8.77E+08\\
WD2105-820	&10559&	8.184&	7.67E+08\\
WD0839-327	&9268	&7.885&	6.67E+08\\
WD0326-273&	9250	&7.86&	6.55E+08\\
WD1544-377&	10538&	8.09	&6.46E+08\\
WD1647+591	&12260&	8.31&	6.06E+08\\
WD2336-079	&11040	& 8.11&	5.89E+08\\
WD1917-077	&10200 &	8&	5.74E+08\\
WD1632+177	&10100	&7.956&	5.67E+08\\
WD1124+595&	10500&	8	&5.31E+08\\
WD2326+049	&11562&	8.008&	4.20E+08\\
WD1935+276	&12130& 	8.05	&4.05E+08\\
WD0806-661	&11940	& 8	&3.79E+08\\
WD2047+372&	14070&	8.21&	3.61E+08\\
WD2341+322&	12570	&7.93	&3.09E+08\\
WD0415-594&	13342&	8	&2.80E+08\\
WD0148+467&	13430&	7.93	&2.56E+08\\
WD1919+145	&15108&	8.078&	2.31E+08\\
WD1327-083&	13920&	7.86	&2.14E+08\\
WD1134+300&	21276	&8.545&	1.93E+08\\
WD2117+539	&13990&	7.78&	1.93E+08\\
WD2007-303&	14454	&7.857&	1.90E+08\\
WD0310-688&	15500&	8.027&	1.90E+08\\
WD0413-077&16176&	7.865	&1.33E+08\\
WD0642-166&	25193&	8.556	&1.12E+08\\
WD0644+375&	21060&	8.1	&8.12E+07\\
WD2032+248	&19980&	7.83	&5.66E+07\\
WD1620-390&	24276&	8.011&	3.01E+07
\enddata
\end{deluxetable}

\begin{deluxetable}{lccrcr}
\tablecaption{Subgroup Velocity Statistics}
\tablewidth{0pc}
\tablenum{5}
\tablehead{
\colhead{Spectral Type}&
\colhead{N}&
\colhead{Component}&
\colhead{Avg.}&
\colhead{Disp.}&
\colhead{Error}
 }  
\startdata                                             
              DA (DA+DAV+DAZ) & 69 &    U &       4.9 &  28.8 &  3.4 \\
                                 &&     V  & -15.8  &  25.5  &  3.0 \\
                                 &&     W   &    -4.3  & 23.1 &    2.7\\
                                  &&    T   &  39.0 &    27.6 &    3.2\\
            MAGNETICS & 17 &    U   & -7.2  &  26.6    & 6.4 \\
                          &&            V    &  -15.7 & 20.1 &    4.8\\
                           &&           W  &   -3.1 &  26.1 &    6.3\\
                            &&          T   &    39.0 & 22.6  & 5.4 \\
                     DC & 21   &  U   &   -8.6 &  53.0  &  11.5 \\
                         &&             V   &   -32.0  &  53.6  &  11.7 \\
                          &&            W   &   -3.6  &  39.7  &   8.6 \\
                          &&            T     &  74.7 &  50.9 &   11.1\\
                     DQ & 12 &    U  &    -4.0 &   42.7  &  12.3\\
                          &&            V   &  -28.5 & 33.9  & 9.7 \\
                           &&           W    & -12.3  &  37.7 & 10.8\\
                            &&          T   & 57.8  & 42.7 &   12.3\\
                    DZ  & 9  &   U  &  20.1 &  39.5   & 13.1\\
                            &&          V     & -34.9  & 42.4 & 14.1\\
                          &&            W      &  -12.0  &  25.5 & 8.5\\
                              &&        T     &  61.8  &  41.3  & 13.7\\   
\enddata
\end{deluxetable}
\newpage             

\section{Figure Captions}

\begin{figure}
\plotone{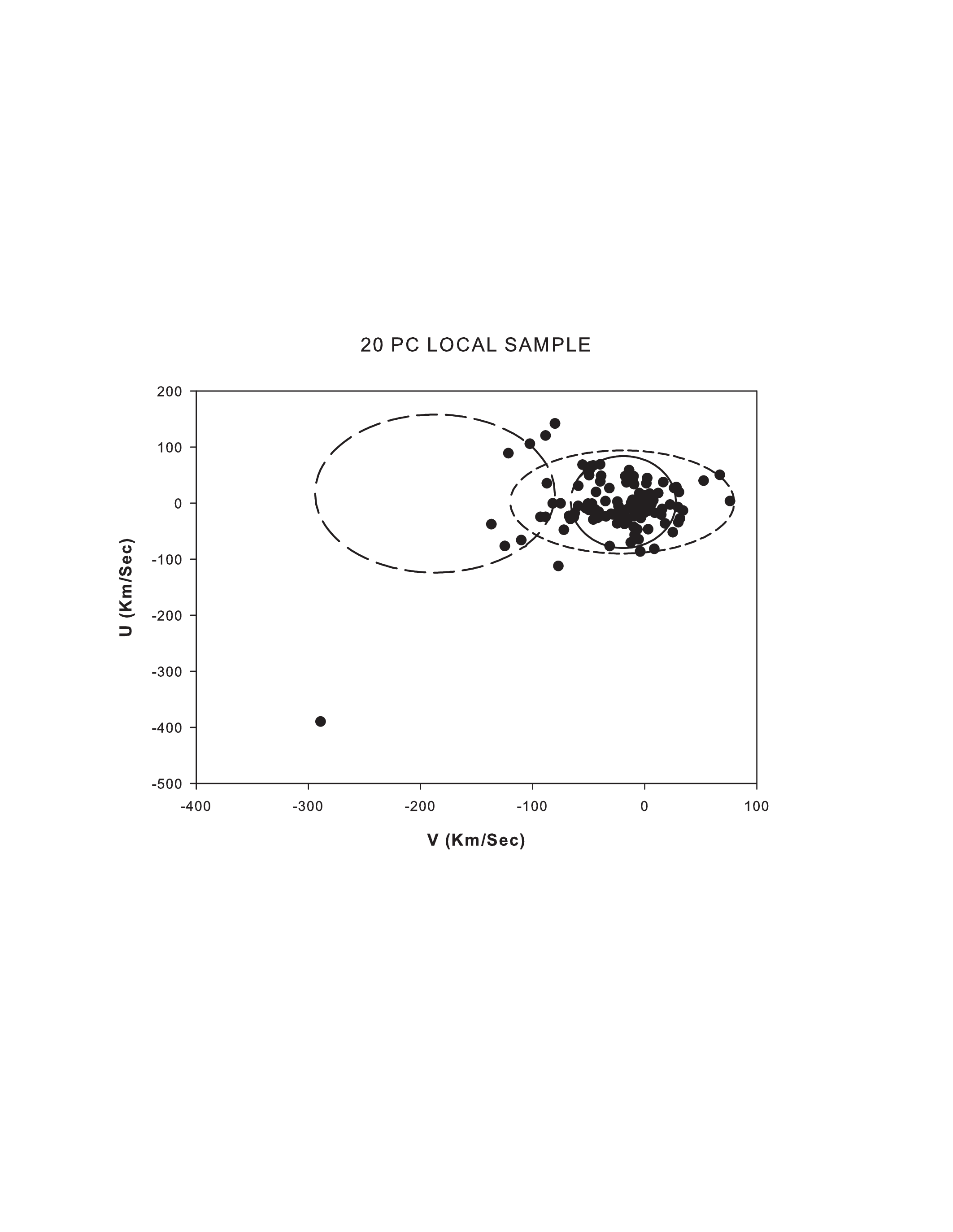}

\caption{The U versus V space velocity diagram for the 20 pc sample of
white dwarfs with the assumption of zero radial velocity. For comparison,
three velocity ellipses for main sequence stars are shown following Chiba
\& Beers (2000; see also Vennes \& Kawka 2006), the $2\sigma$ velocity
ellipse contour (solid line) of the thin disk component, the $2\sigma$
ellipse of the thick disk component (short-dashed line)  and the $1\sigma$
contour of the halo component (long-dashed line).}

\end{figure}

\end{document}